\documentclass[useAMS,usenatbib]{mn2e}
\usepackage{multirow}
\usepackage{epsfig}
\usepackage{subfigure}
\usepackage[T1, hyphens]{url}
\usepackage{tikz}
\usepackage{pifont}
\usepackage{wasysym}
\usepackage{amssymb}
\makeatletter
\let\jnl@style=\rm
\def\ref@jnl#1{{\jnl@style#1}}

\def\aj{\ref@jnl{AJ}}                   % Astronomical Journal
\def\araa{\ref@jnl{ARA\&A}}             % Annual Review of Astron and Astrophys
\def\apj{\ref@jnl{ApJ}}                 % Astrophysical Journal
\def\apjl{\ref@jnl{ApJ}}                % Astrophysical Journal, Letters
\def\apjs{\ref@jnl{ApJS}}               % Astrophysical Journal, Supplement
\def\ao{\ref@jnl{Appl.~Opt.}}           % Applied Optics
\def\apss{\ref@jnl{Ap\&SS}}             % Astrophysics and Space Science
\def\aap{\ref@jnl{A\&A}}                % Astronomy and Astrophysics
\def\aapr{\ref@jnl{A\&A~Rev.}}          % Astronomy and Astrophysics Reviews
\def\aaps{\ref@jnl{A\&AS}}              % Astronomy and Astrophysics, Supplement
\def\azh{\ref@jnl{AZh}}                 % Astronomicheskii Zhurnal
\def\baas{\ref@jnl{BAAS}}               % Bulletin of the AAS
\def\jrasc{\ref@jnl{JRASC}}             % Journal of the RAS of Canada
\def\memras{\ref@jnl{MmRAS}}            % Memoirs of the RAS
\def\mnras{\ref@jnl{MNRAS}}             % Monthly Notices of the RAS
\def\pra{\ref@jnl{Phys.~Rev.~A}}        % Physical Review A: General Physics
\def\prb{\ref@jnl{Phys.~Rev.~B}}        % Physical Review B: Solid State
\def\prc{\ref@jnl{Phys.~Rev.~C}}        % Physical Review C
\def\prd{\ref@jnl{Phys.~Rev.~D}}        % Physical Review D
\def\pre{\ref@jnl{Phys.~Rev.~E}}        % Physical Review E
\def\prl{\ref@jnl{Phys.~Rev.~Lett.}}    % Physical Review Letters
\def\pasp{\ref@jnl{PASP}}               % Publications of the ASP
\def\pasj{\ref@jnl{PASJ}}               % Publications of the ASJ
\def\qjras{\ref@jnl{QJRAS}}             % Quarterly Journal of the RAS
\def\skytel{\ref@jnl{S\&T}}             % Sky and Telescope
\def\solphys{\ref@jnl{Sol.~Phys.}}      % Solar Physics
\def\sovast{\ref@jnl{Soviet~Ast.}}      % Soviet Astronomy
\def\ssr{\ref@jnl{Space~Sci.~Rev.}}     % Space Science Reviews
\def\zap{\ref@jnl{ZAp}}                 % Zeitschrift fuer Astrophysik
\def\nat{\ref@jnl{Nature}}              % Nature
\def\iaucirc{\ref@jnl{IAU~Circ.}}       % IAU Cirulars
\def\aplett{\ref@jnl{Astrophys.~Lett.}} % Astrophysics Letters
\def\apspr{\ref@jnl{Astrophys.~Space~Phys.~Res.}}
\def\bain{\ref@jnl{Bull.~Astron.~Inst.~Netherlands}}
\def\fcp{\ref@jnl{Fund.~Cosmic~Phys.}}  % Fundamental Cosmic Physics
\def\gca{\ref@jnl{Geochim.~Cosmochim.~Acta}}   % Geochimica Cosmochimica Acta
\def\grl{\ref@jnl{Geophys.~Res.~Lett.}} % Geophysics Research Letters
\def\jcp{\ref@jnl{J.~Chem.~Phys.}}      % Journal of Chemical Physics
\def\jgr{\ref@jnl{J.~Geophys.~Res.}}    % Journal of Geophysics Research
\def\jqsrt{\ref@jnl{J.~Quant.~Spec.~Radiat.~Transf.}}
\def\memsai{\ref@jnl{Mem.~Soc.~Astron.~Italiana}}
\def\nphysa{\ref@jnl{Nucl.~Phys.~A}}   % Nuclear Physics A
\def\physrep{\ref@jnl{Phys.~Rep.}}   % Physics Reports
\def\physscr{\ref@jnl{Phys.~Scr}}   % Physica Scripta
\def\planss{\ref@jnl{Planet.~Space~Sci.}}   % Planetary Space Science
\def\procspie{\ref@jnl{Proc.~SPIE}}   % Proceedings of the SPIE

\makeatother

\newcommand{\xmark}{\ding{55}}%
\newcommand {\apgt} {\ {\raise-.5ex\hbox{$\buildrel>\over\sim$}}\ }
\newcommand {\aplt} {\ {\raise-.5ex\hbox{$\buildrel<\over\sim$}}\ } 

\title[Spatially resolved Fe K spectroscopy of NGC 4945]{Spatially resolved Fe K spectroscopy of NGC 4945 }
\author[Andrea Marinucci, et al.]{A. Marinucci$^{1}$\thanks{E-mail: marinucci@fis.uniroma3.it (AM)}, S. Bianchi$^{1}$, G. Fabbiano$^{2}$, G. Matt$^{1}$, G. Risaliti$^{3}$,   
\newauthor
E. Nardini$^{3}$, J. Wang$^{4}$\\
\\
$^1$Dipartimento di Matematica e Fisica, Universit\`a degli Studi Roma Tre, via della Vasca Navale 84, 00146 Roma, Italy\\
$^2$Harvard-Smithsonian Center for Astrophysics, 60 Garden Street, Cambridge, MA 02138, USA\\
$^3$Istituto Nazionale di Astrofisica (INAF) – Osservatorio Astrofisico di Arcetri, Largo Enrico Fermi 5, 50125 Firenze, Italy\\
$^{4}$Department of Astronomy and Institute of Theoretical Physics and Astrophysics, Xiamen University, Xiamen, China 361005\\
}
\begin{document}
\maketitle
\label{firstpage}
\begin{abstract} We present the imaging and spectroscopic analysis of the combined {\it Chandra} ACIS-S observations of the Compton-thick Seyfert 2 galaxy NGC 4945. We performed a spatially-resolved spectroscopy of the circumnuclear environment of the source, picturing the innermost 200 parsecs around the highly absorbed nucleus. The additional 200 ks ACIS-S data with respect to the previous campaign allowed us to map with even greater detail the central structure of this source and to discover an enhanced iron emission in the innermost nuclear region, with respect to the associated Compton reflection continuum. We revealed that the Equivalent Width of the iron K$\alpha$ line is spatially variable (ranging from 0.5 to 3 keV), on scales of tens of parsecs, likely due to the ionization state and orientation effects of the reprocessing material, with respect to the central X-ray illuminating source. A clump of highly ionized Fe \textsc{xxv} He-$\alpha$ is also detected, 40 parsecs east to the nucleus. When observations taken years apart are considered, the central unresolved reflected emission is found to remain constant. 

\end{abstract}

\section{Introduction}
The circumnuclear environment of Active Galactic Nuclei (AGN) is very complex, with several physical components (broad emission line clouds, a dusty ``torus'', a low density ionized gas) contributing to the observed observational properties \citep[e.g.][]{n15}.
Among AGN, Compton-thick sources, characterized by a column density of gas in excess of the inverse of the Thomson scattering cross-section (N$_{\rm H}=\sigma_T^{-1}\simeq 1.25\times10^{24}$ cm$^{-2}$) along the line of sight, are quite common in the local Universe \citep[about 20--30\%,][]{bag11,ruk15}. However, we have little or no observational information on their nuclear structure, due to the thick screen preventing a direct view of the inner regions on the high-frequency part of the electromagnetic spectrum, with the possible exception of the hard ($>$10 keV) X-rays.\\
On the other hand, the spectra of Compton-thick Seyfert 2 galaxies are dominated by reflection components in the X-ray band, from both cold and ionized circumnuclear material \citep{matt00}. 
The heavy obscuration of the primary radiation from the nucleus, at least up to 10 keV, permits a clear view of these components, which are strongly diluted in unobscured sources. Indeed, most of our knowledge about the circumnuclear reprocessing matter, at least as far as their X-ray properties are concerned, is based on the brightest Compton-thick sources, like NGC 1068 \citep{kin02,matt04,baw15, mbm16}, Circinus \citep[e.g.][]{matt96,mbm03,arevalo14} and Mrk 3 \citep[e.g.][]{bianchi05b,pp05,glm12,gra16}.

 \begin{figure*}
 \epsfig{file=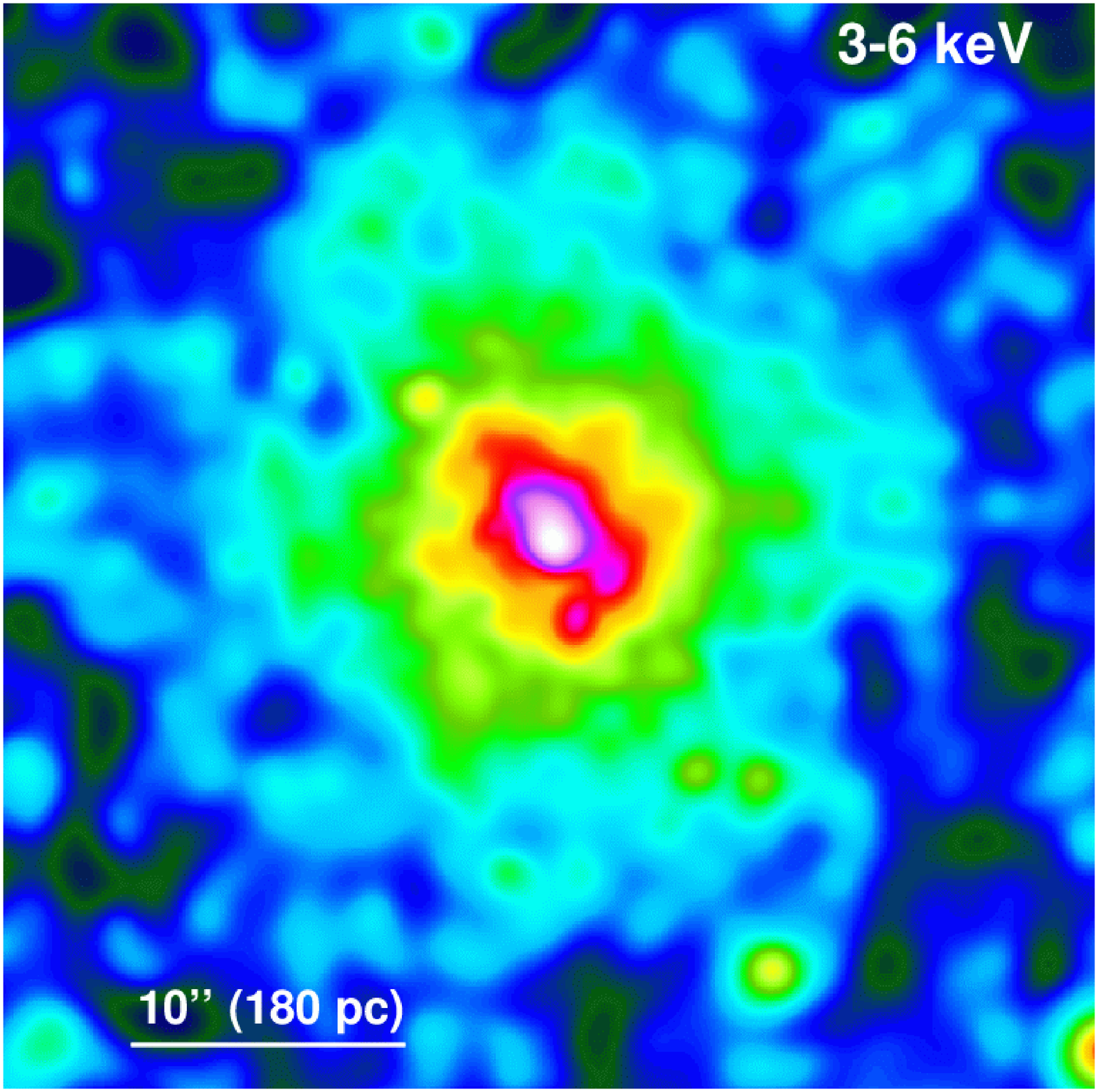, width=0.68\columnwidth}
\epsfig{file=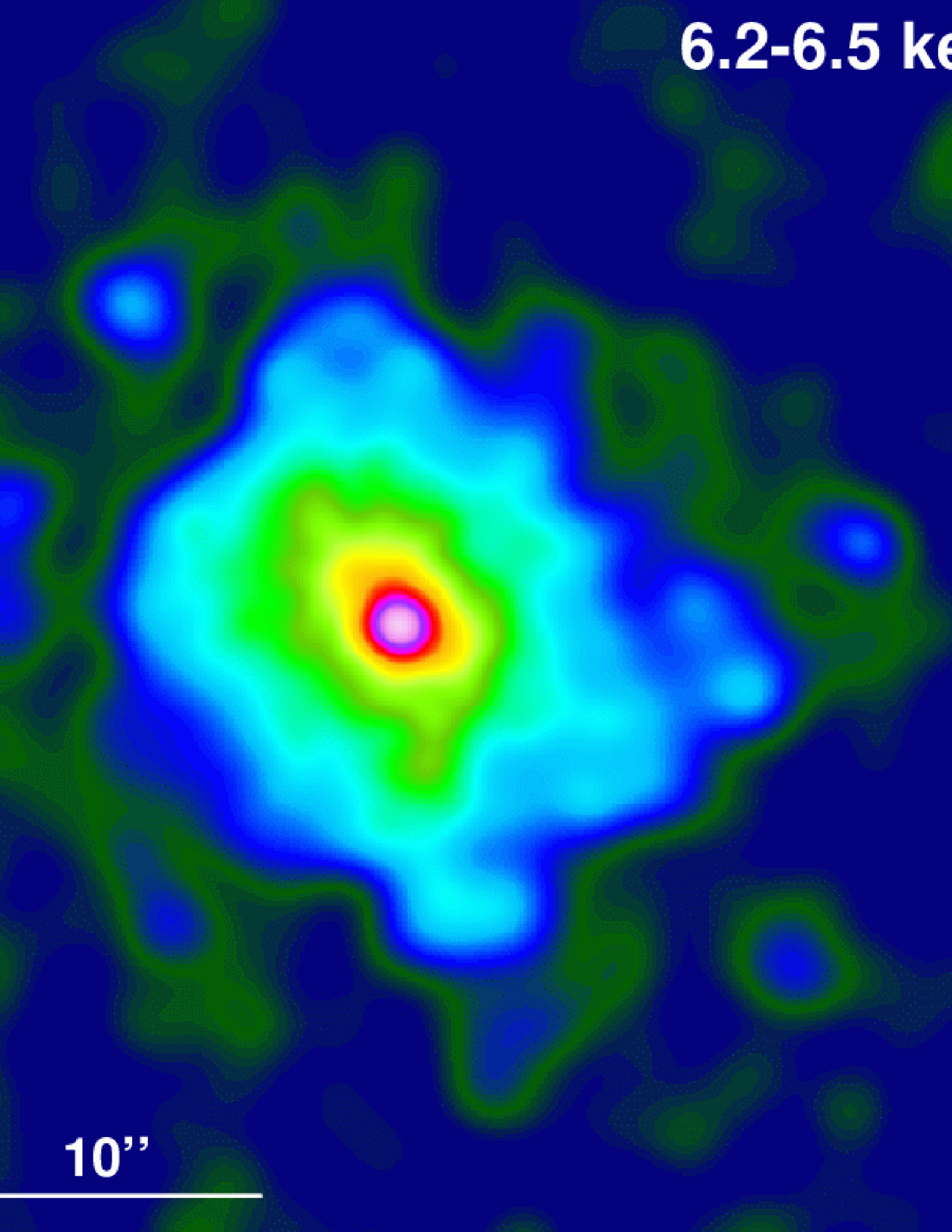, width=0.675\columnwidth}
\epsfig{file=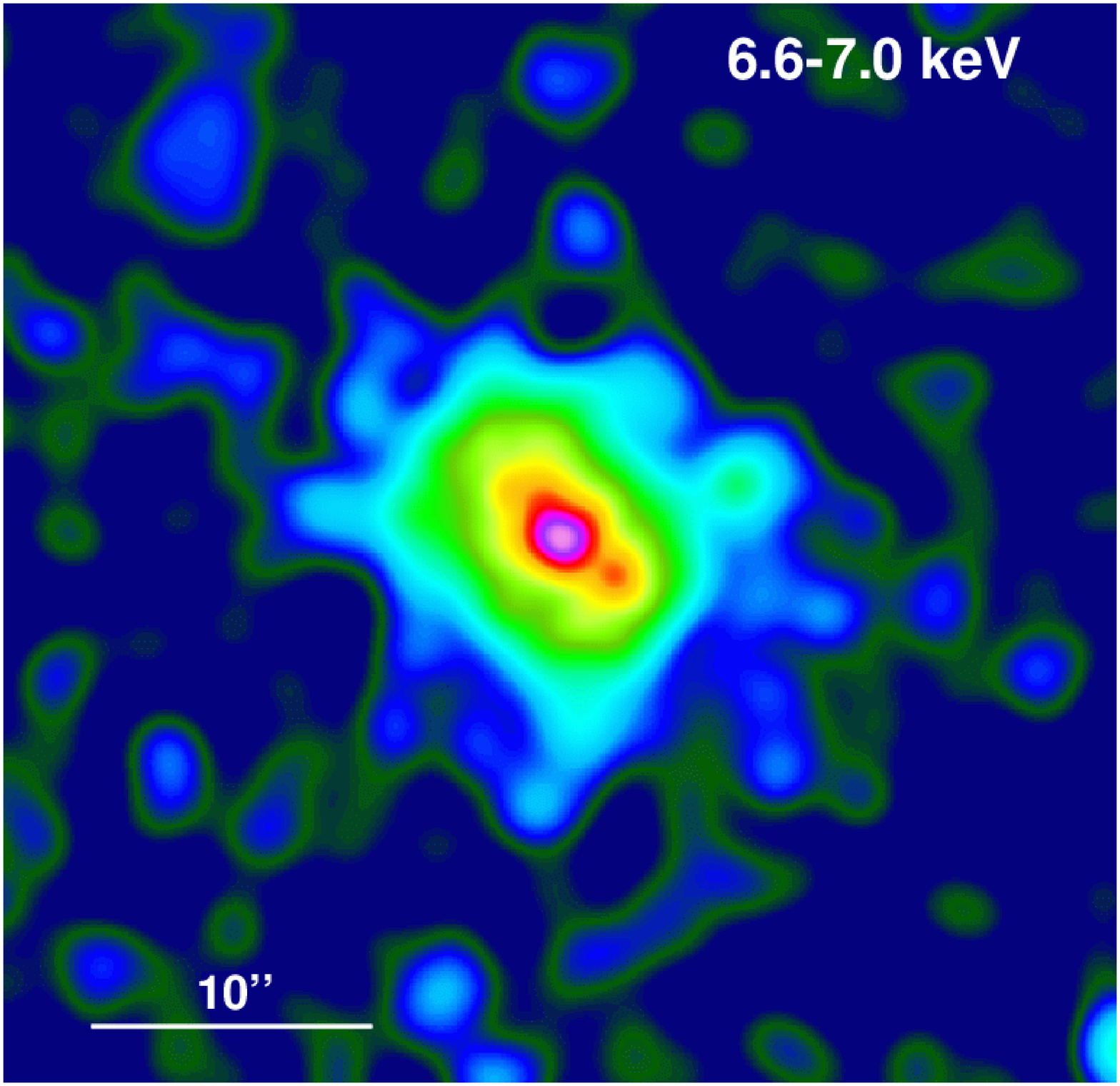, width=0.7\columnwidth}
  \caption{\textit{Left panel:} Merged ACIS-S image of the central 40$\times$40 arcsec region of NGC 4945, in the 3--6 keV energy band. \textit{Central panel:} Merged ACIS-S image of the central 40$\times$40 arcsec region of NGC 4945, in the 6.2--6.5 keV energy band. \textit{Right panel:} Merged ACIS-S image of the central 40$\times$40 arcsec region of NGC 4945, in the 6.6--7.0 keV energy band. We adopted a subpixeling factor of eight (the size of a pixel is therefore 0.062 arcsec) in all the images. Adaptive smoothing is performed with Gaussian kernels from 0.5 to 15 pixels, 30 log steps.}
  \label{subpix_images}
\end{figure*}

 NGC 4945 is one of the nearest AGN (D $\sim$3.7~Mpc) and the brightest Seyfert 2 in the 100 keV sky \citep{dms96}, with a highly absorbed \citep[$N_H=3.5\times 10^{24}$ cm$^{-2}$,][]{pcf14} and extremely variable primary continuum \citep{gmb00,mzd00, dmz03}. Such a strong absorption along the line of sight completely blocks the nuclear emission below 10 keV and allows it to be observed only at higher energies, though heavily affected by Compton scattering and photoelectric absorption. 
Broadband analyses of its high energy variability \citep{yaq12, pcf14} led to important constraints on the geometry of the reflector: the  variability above 10~keV, together with a constant reflection  suggests that most of the observed emission is due to the intrinsic continuum, and not to Compton scattering from other directions. This suggests that the reflector covers a very small solid angle $<10^{\circ}$ as seen from the source (assuming a toroidal structure).  This physical scenario is in agreement with the very low ratio between the reflected emission below 10~keV and the intrinsic flux in the same band as estimated from the emission above 10~keV \citep[$<$0.1\%,][]{dmz03}. 
 \citet{mrw12} compared the strong intrinsic variability measured by {\em Swift/BAT} to the constant reflection spectra from {\em XMM-Newton} and {\em Suzaku} observations over more than ten years, finding a distance of the reflecting structure D$>$35~pc. {\em Chandra} is the only present X-ray instrument able to resolve this angular dimension and indeed an extended clumpy structure on projected scales of $\sim$200$\times$100~pc has been revealed. When the images of the reflection continuum and the associated Iron K$\alpha$ emission line are compared (extracted in the 2--10 keV and 6.2--6.7 keV bands, respectively) it has been shown that the emitting material is one and the same, extended well beyond the nuclear region. However, past archival data (before the additional 2013 observing campaign) do not allow us to spectrally  characterize different extra-nuclear regions with high detail, due to low signal-to-noise, nor to investigate possible ionization structures of the Fe emitting material. \\
 We hereby analyse and discuss recent {\it Chandra} ACIS-S observations of NGC~4945 which, combined with archival data, lead to the detection of spatially resolved neutral and ionized Fe emitting material, on scales of hundreds of parsecs.  The paper is structured as follows: in Sect. 2 we discuss the combined {\it Chandra} observations and data reduction, in Sect. 3  we present the imaging and spectral analyses, respectively. We discuss and summarize the physical implications of our results in Sect. 4 and 5. At the distance of the source 1 arcsec=18 pc.\\
 
 \begin{figure*}
\epsfig{file=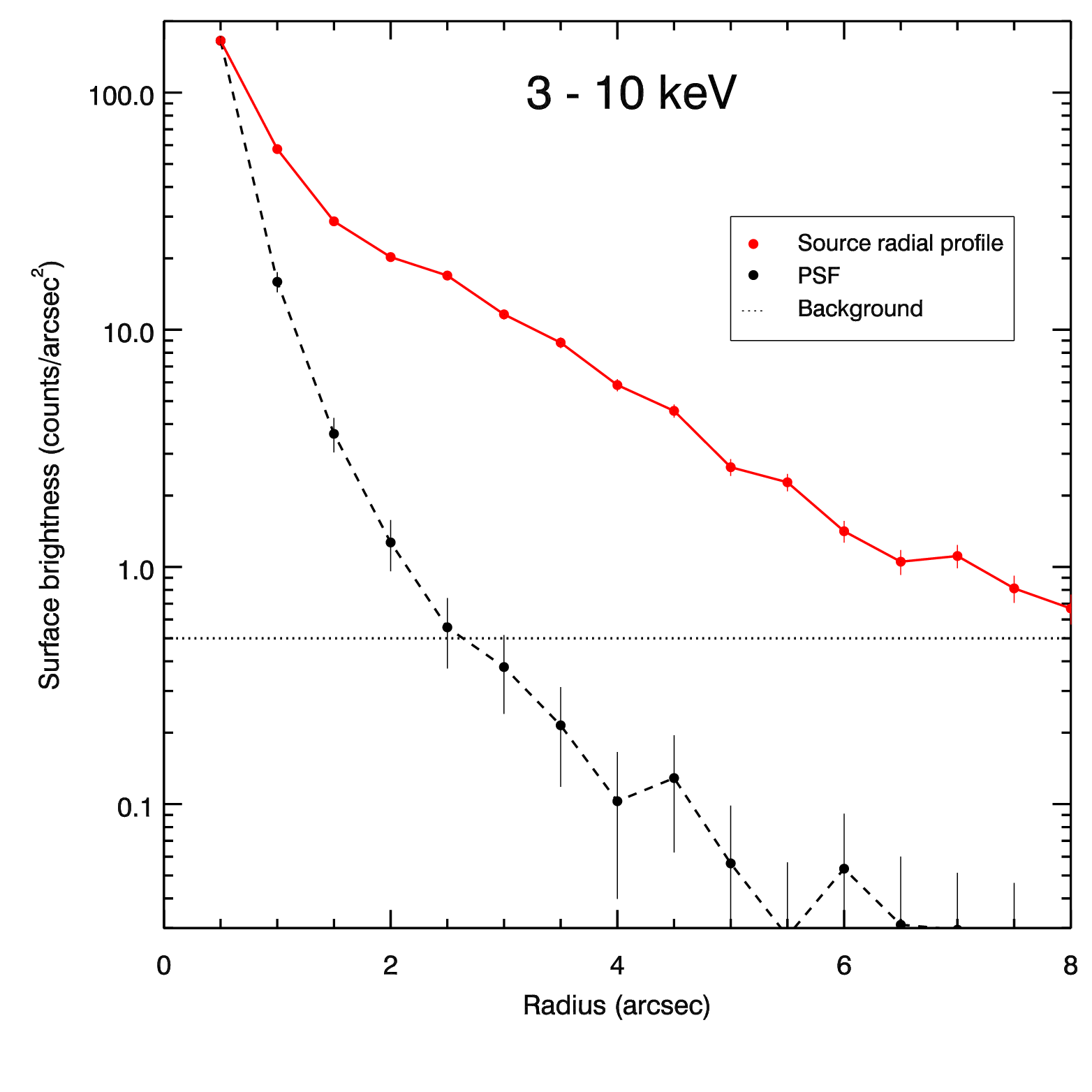, width=0.68\columnwidth}
\epsfig{file=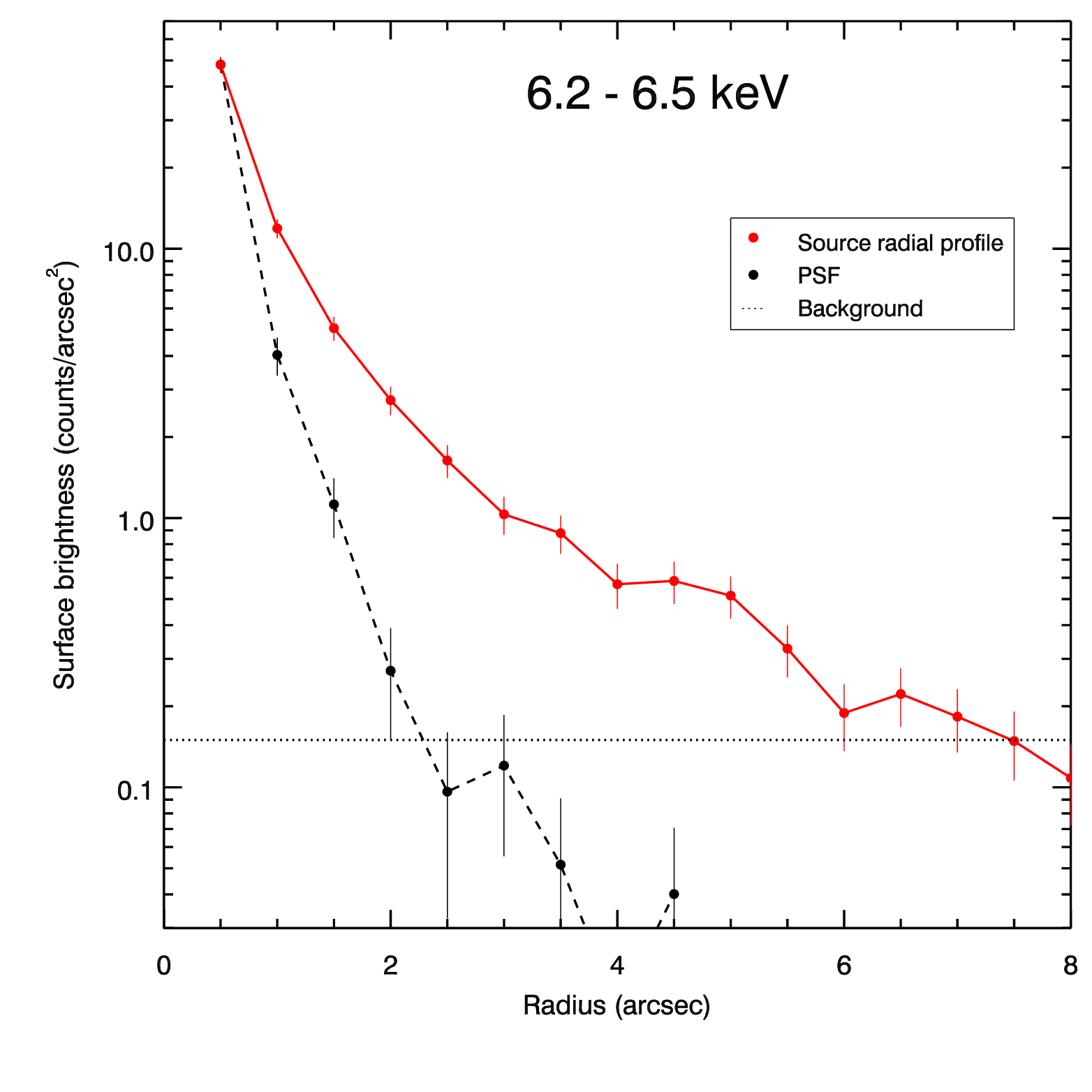, width=0.68\columnwidth}
\epsfig{file=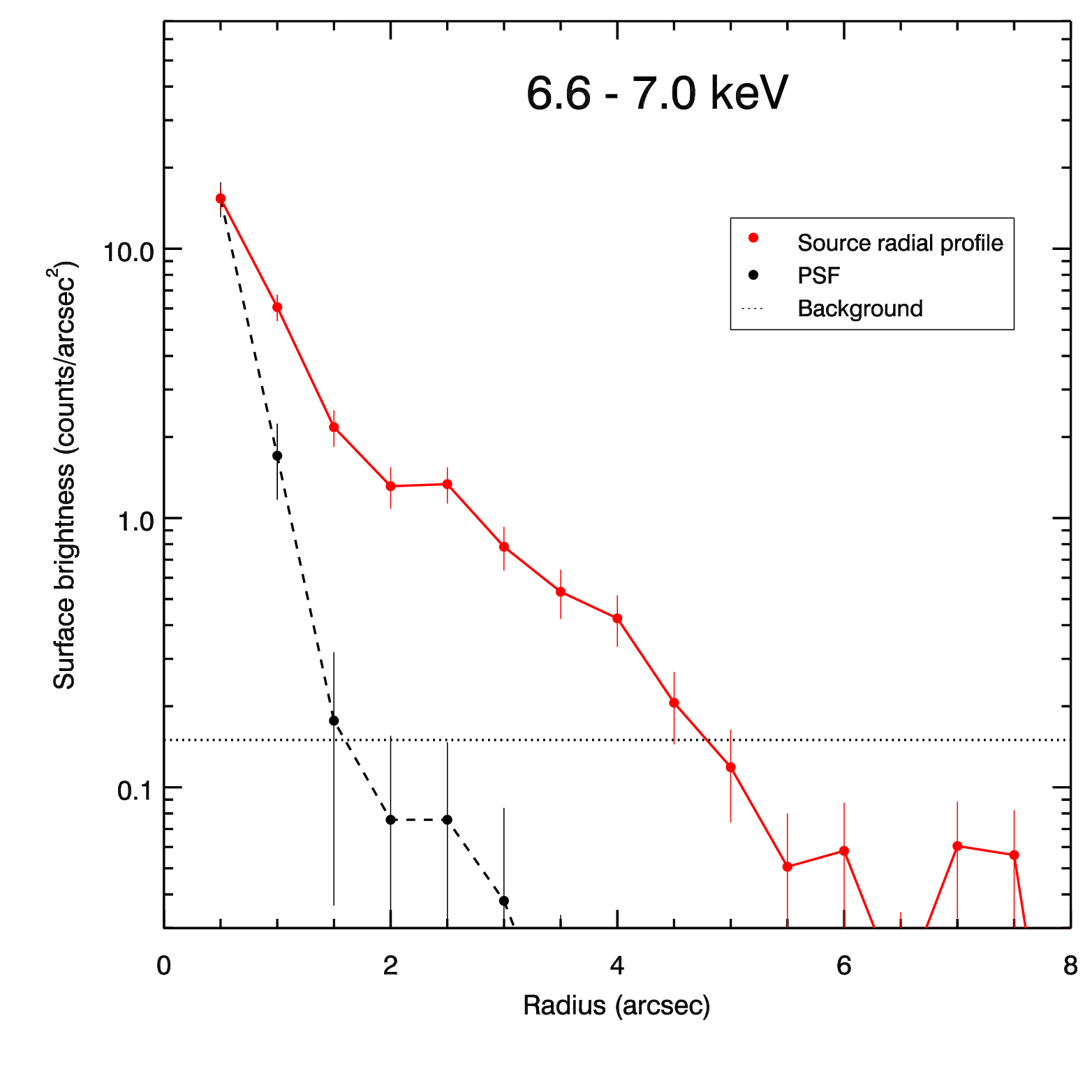, width=0.68\columnwidth}
  \caption{\textit{Left panel:} Radial profiles of the source and ACIS-S Point Spread Function in the total 3-10 keV energy band (lines and continuum). \textit{Central panel:}  Radial profiles of the source and ACIS-S Point Spread Function in 6.2-6.5 keV energy band.\textit{Right panel:}  Radial profiles of the source and ACIS-S Point Spread Function in 6.6-7.0 keV energy band. We chose 1 native pixel radius, concentric annuli for the source and PSF surface brightness, and a 25 pixel radius annulus for the background one. The PSF radial profile was normalized to the one of the source.  Surface brightness of the background was inferred using a source free region, in the southwest direction from the nucleus, distant from the diffuse emission.}
  \label{radprof}
\end{figure*}

\begin{table}
\begin{center}
\begin{tabular}{cccc}
{\bf Obs. ID} & {\bf Date} & {\bf Exp. Time (ks)} & {\bf HETG } \\
\hline
864 & 2000-01-27    &  49.7& \xmark \\
4899  &   2004-05-28  &    78.6 &  \checkmark \\
4900   &  2004-05-29   &   95.8  &  \checkmark\\
14985  & 2013-04-20    &  68.7&\xmark \\
14984  & 2013-04-25   & 130.5&\xmark \\
\hline
\end{tabular}
\end{center}
\caption{\label{chaobs} Observation log for the {\it Chandra} ACIS-S observations of NGC 4945.}
\end{table}

\section{Observations and data reduction}
NGC 4945 \citep[z=0.001878,][]{trs09} has been intensively observed by {\it Chandra} with the Advanced CCD Imaging Spectrometer \citep[ACIS:][]{acis} in the last few years. We report the five observations of interest in Table \ref{chaobs}. We excluded ObsIds. 13791 and 14412 on purpose, since our target is strongly off-axis ($\sim 4.28$ arcmin) and therefore the PSF is heavily degraded. Data were reduced with the Chandra Interactive Analysis of Observations \citep[CIAO:][]{ciao} 4.8 and the Chandra Calibration Data Base (CALDB) 4.7.2 database, adopting standard procedures. We generated event files for the five observations separately with the CIAO tool \textsc{chandra\_repro} and merged them into a single image using the tool \textsc{merge\_all}. After cleaning for background flaring we got a total exposure of 420 ks. The imaging analysis was performed applying the Subpixel Event Repositioning and adaptive smoothing procedures widely discussed in the literature \citep[][and references therein]{tsunemi01, li04, wang11b, mfe16}. \\
Spectra from the nucleus of the source were extracted from the central circular region, adopting a 1 arcsec radius, in all the five observations. We used a $5$ arcsec radius circle for background extraction located in a source free region, in the southwest direction from the nucleus, well away from the diffuse emission. Spectra were binned in order to over-sample the instrumental resolution by a factor of 3 and to have no less than 25~counts in each background-subtracted spectral channel. This allows the applicability of the $\chi^2$ statistics. \\
Spectra from regions 1, 2 and 3 were extracted from 1 arcsec circular regions while we used a 2 arcsec radius for region 4, from the five different observations ( Fig. 3, more details on the extraction regions are given in Section 3). We then co-added spectra extracted from ObsIds. 864, 14984, 14985, and spectra from ObsIds. 4899 and 4900, to maximize the S/N ratio. Due to the presence of the HETG in ObsIds. 4899 and 4900, the effective collecting area is reduced by $\sim40\%$ at 6 keV with respect to ObsIds. 864, 14984, 14985.  Due to the low count rates we binned the spectra using 10 counts per channel and used the Cash statistics \citep[C-stat:][]{cash76} in our fitting procedure. Total count rates are summarized in Table \ref{chacounts}.\\
Pileup is constrained to be lower than 1\% \citep[following the procedure described in][]{davis01}.
We adopt the cosmological parameters $H_0=70$ km s$^{-1}$ Mpc$^{-1}$, $\Omega_\Lambda=0.73$ and $\Omega_m=0.27$, i.e. the default ones in \textsc{xspec 12.9.0} \citep{xspec}. Errors correspond to the 90\% confidence level for one interesting parameter ($\Delta\chi^2=2.7$), if not stated otherwise. 

\begin{table}
\begin{center}
\begin{tabular}{cccccc}
{\bf Obs. ID} &  \multicolumn{4}{c}{\bf 3--10 keV Counts}\\
\hline
 &  \multicolumn{4}{c}{Extraction region} \\
 &  1 & 2 & 3 & 4 & Nucleus \\
%  &   &  &  &  \\
  \hline
864$\ $ \multirow{3}{*}{\Bigg{\}}}&     &  &  & &359\\
14984  $\ \ \ $&    361 & 303 & 460& 148 &1029\\
14985  $\ \ \ $&    & & & &588\\
4899 \multirow{2}{*}{\big{\}}} & \multirow{2}{*}{116}    &  \multirow{2}{*}{138}   & \multirow{2}{*}{163} & \multirow{2}{*}{46} &370\\
4900  $\ \ \ $ &    &    &  & & 424\\
\hline
\end{tabular}
\end{center}
\caption{\label{chacounts} Number of 3--10 keV counts from different extraction regions are reported. We refer to Fig. \ref{images} (top-right and bottom-left panels) for the exact spatial locations.}
\end{table}
\begin{figure*}
 \epsfig{file=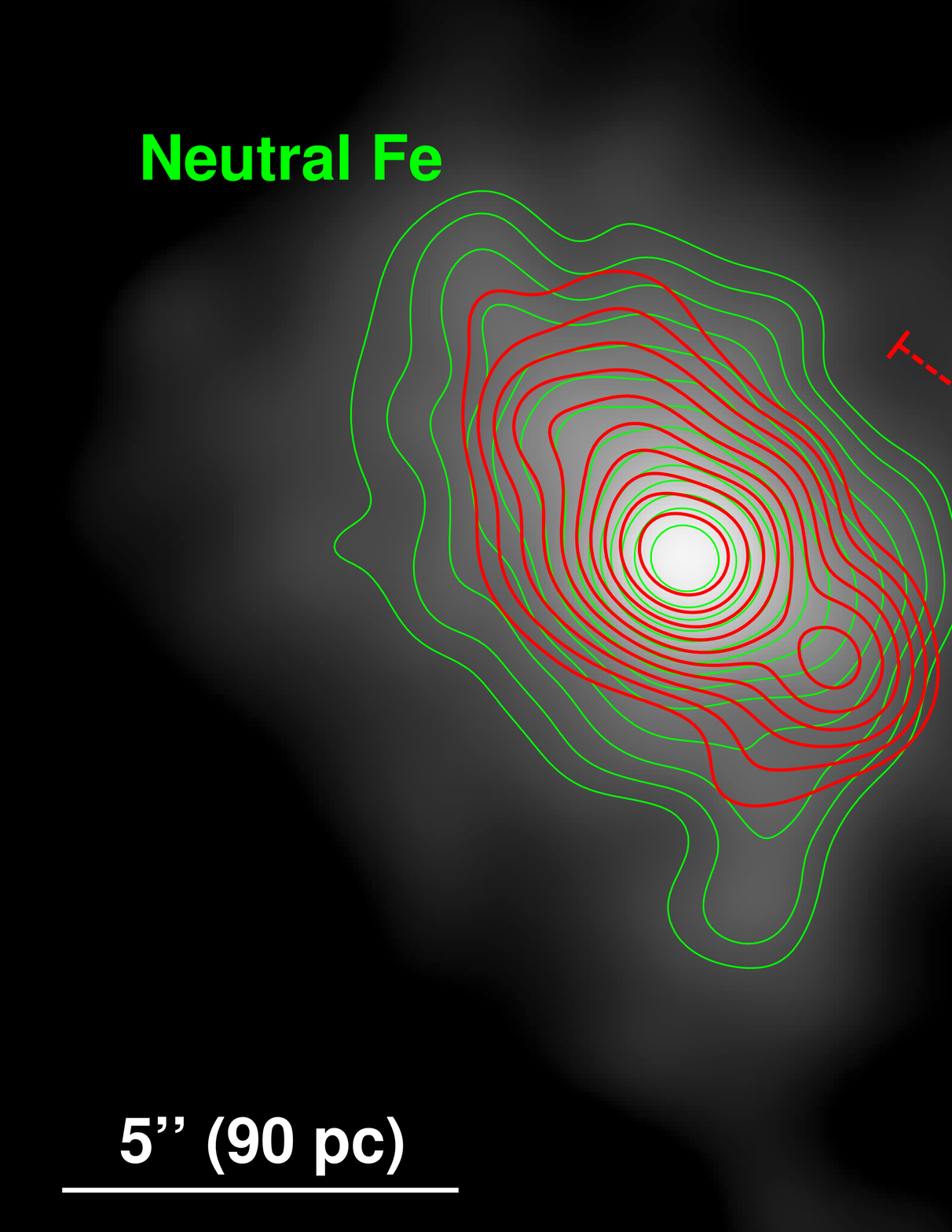, width=8.2cm}
  \epsfig{file=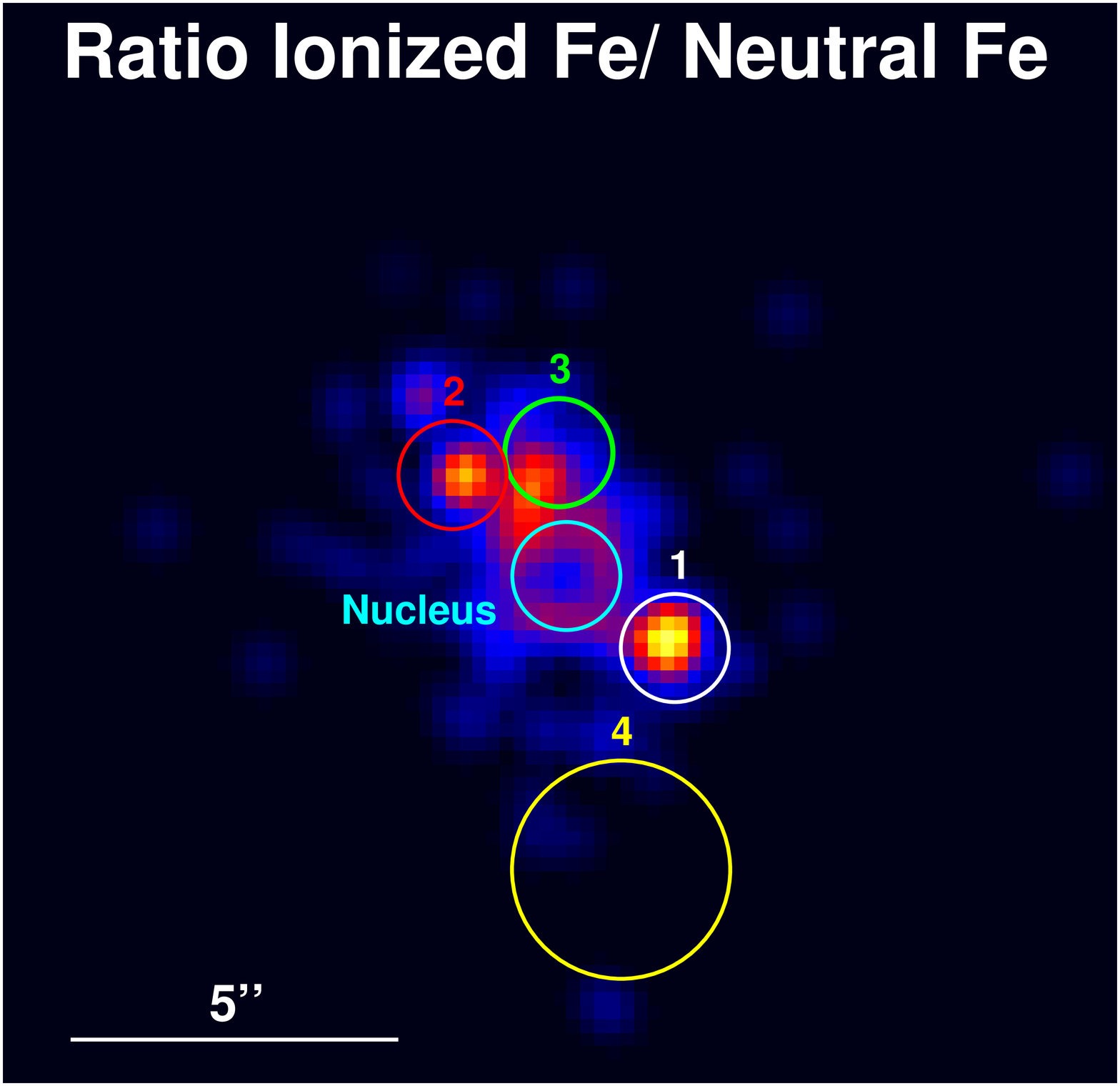, width=8.2cm, height=8.25cm}\\
 \epsfig{file=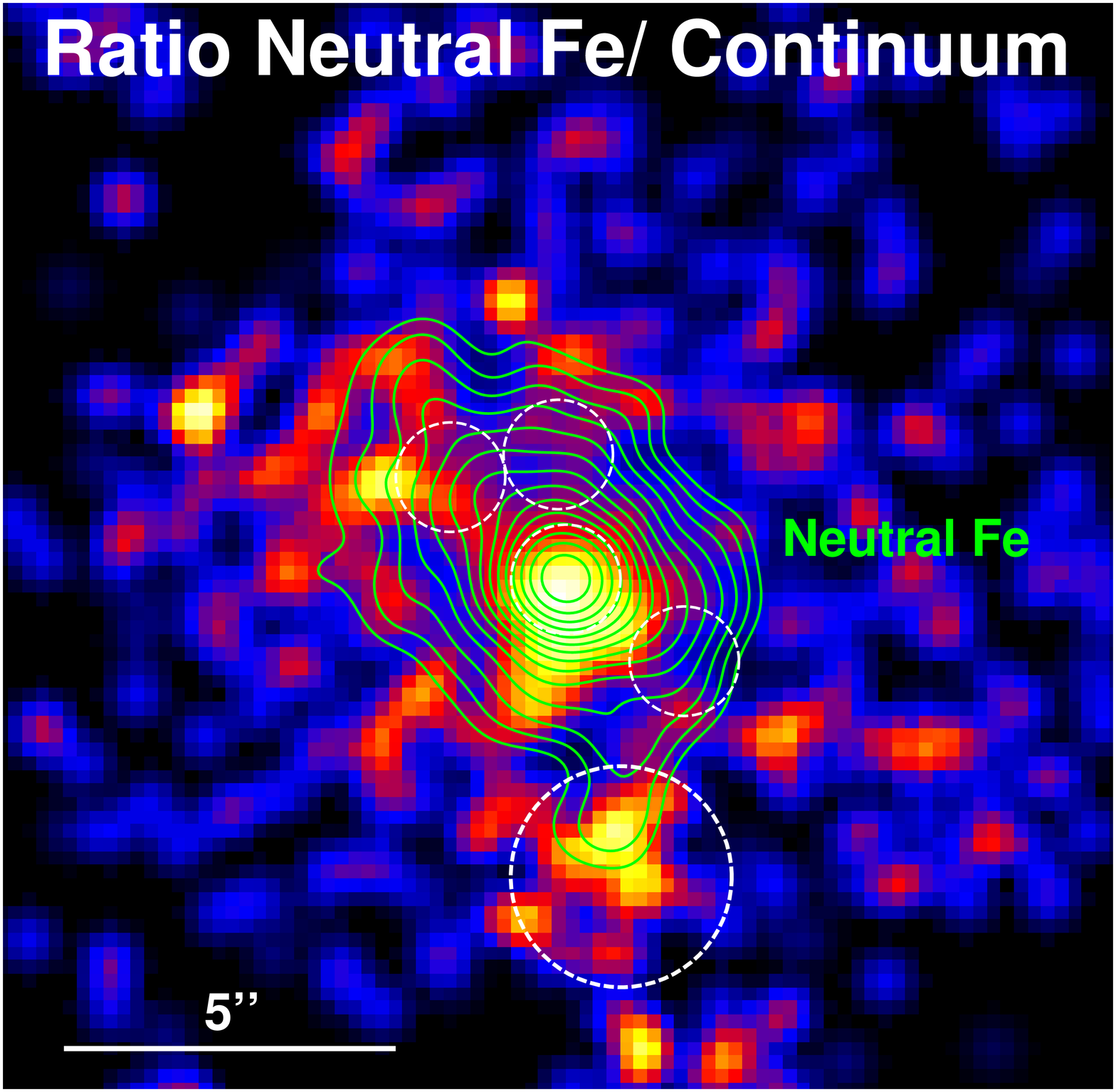, width=8.2cm}
 \epsfig{file=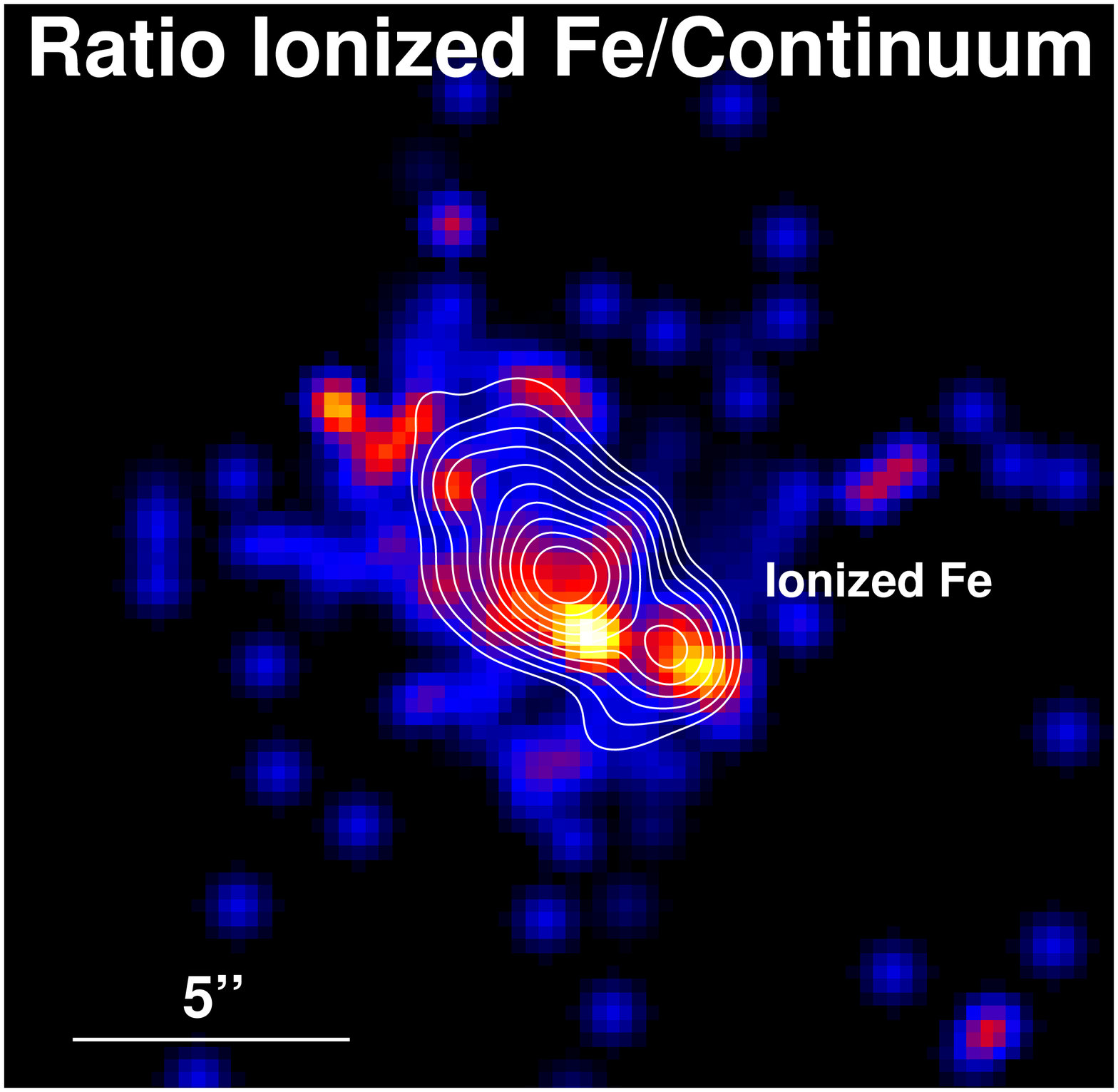, width=8.2cm, height=8.02cm}
  \caption{\textit{Top-left panel:} Solid green and red lines indicate iso-counts contours from the 6.2--6.5 keV and 6.6--7.0 keV images, respectively. \textit{Top-right panel:} Ratio between the 6.6--7.0 keV and 6.2--6.5 keV images, with over-imposed circular extraction regions: 1 arcsec radii have been adopted for all regions with the exception of region 4. \textit{Bottom-left panel:}  Ratio between the iron K$\alpha$ and continuum only images, with contours extracted from the 6.2--6.5 keV image. The peak of the ratio well resembles the structure of the iron K$\alpha$ emitting material. Extraction regions are shown as dashed circles. \textit{Bottom-right panel:} Ratio between the ionized iron and continuum only images, with contours extracted from the 6.6--7.0 keV image. We adopted a subpixeling factor of a half (the size of a pixel is therefore 0.247 arcsec) and a Gaussian smoothing with a 3 pixels kernel in all the images.}
  \label{images}
\end{figure*}
\section{Data analysis}
\subsection{Imaging}
Throughout this paper, we focus our analysis to the energy band above 3 keV, to eliminate any contribution from the diffuse, soft emitting gas, firstly reported in \citet{srw02}. We show, in Fig. \ref{subpix_images}, the 40 arcsec $\times$ 40 arcsec central region. The 420 ks long merged event file was filtered in energy between 3-6 keV (left panel), 6.2-6.5 keV (middle panel) and 6.6-7.0 keV (right panel). The first band was chosen to image the circumnuclear material responsible for the Compton reflection continuum excluding any lines contribution while the other two bands were chosen to map the neutral Fe emission and the ionized (mainly Fe \textsc{xxv} He-$\alpha$ and Fe \textsc{xxvi} Ly-$\alpha$) iron lines, respectively. We then simulated, using the X-ray tracing code \textsc{marx} \citep{dbd12}, the ACIS-S PSF contribution. The peak of the PSF emission is normalized to the peak of the source radial profiles. We show, in Fig. \ref{radprof}, the radial profiles in three energy bands: the source emission is extended and well above the PSF contribution. Indeed,  emission from Fe \textsc{xxv} He-$\alpha$ and Fe {xxvi} Ly-$\alpha$ lines is also extended and shows a  particularly bright clump $\sim2$ arcsec far to the right of the nucleus (see Fig. \ref{subpix_images}, right panel).\\
\noindent
Fig. \ref{subpix_images} also clearly shows that iron K$\alpha$, iron \textsc{xxv} He-$\alpha$ and iron {xxvi} Ly-$\alpha$ emitting materials are not exactly distributed on the same spatial scale. We therefore extracted iso-counts contours from the 6.2--6.5 keV and 6.6--7.0 keV images and over-imposed them on the 3--6 keV image (Fig. \ref{images}, top-left panel: green and red solid lines, respectively). While the peak of the two emission lines is confined towards the nucleus, a bright ionized iron clump emerges at 2.2 arcsec from the nuclear peaks, corresponding to a distance d=40 parsecs at the redshift of NGC 4945.
Some further considerations can be drawn if we divide the 6.2--6.5 keV image by the 3--6 keV one. This gives us indications about the strength of the iron K$\alpha$ emission line with respect to the associated reflection continuum (a proxy for the EW) which is seen to be enhanced in the nucleus and in a southern clump (Fig. \ref{images}, bottom-left panel). On the other hand, when we consider the ratio between the 6.6--7.0 keV and 6.2--6.5 keV images the bright ionized iron clump is clearly visible 2.2 arcsec far from the nucleus (Fig. \ref{images}, top-right panel). Also the ratio between the 6.6--7.0 keV and the 3--6 keV image shows an extended material coincident with the ionized Fe contours (Fig. \ref{images}, bottom-right panel). We chose five different extraction regions, which are shown in Fig. \ref{images}. The nuclear one is centered on the nucleus of the source, region 1 is centered on the ionized iron clump, region 2 on a region which is symmetric to the clump with respect to the nucleus and region 3 on a generic region close to the nucleus but on a different line direction with respect to the former regions, far from the core of the diffuse soft emission. Region 4 is chosen to match the southern clump which is visible in the ratio between the neutral Fe and the continuum.

\begin{table*}
\begin{center}
\begin{tabular}{ccccccc}
{\bf Parameter} &  {\bf All-tied} & {\bf Obs. 1} & {\bf Obs. 2} &{\bf Obs. 3} & {\bf Obs. 4} & {\bf Obs. 5} \\
\hline
N$_{\rm pexrav}$ & $1.55\pm0.10$& $1.80\pm0.20$ &$1.90\pm0.30$ & $1.60\pm0.20$&$1.40\pm0.10$&$1.55\pm0.15$\\
\hline
${{\rm Fe\ K} \alpha}$ & & & & & \\
Energy&$6.414\pm0.007$& -&-& -&-& - \\
Flux &$1.25\pm0.10$ & $1.00^{+0.15}_{-0.25}$&$1.20\pm0.30$ & $1.40\pm0.30$& $1.35^{+0.20}_{-0.15}$& $1.40\pm0.20$\\
EW & $2.7\pm0.5$&$1.9^{+0.5}_{-0.7}$ & $2.0\pm0.6$&$3.0\pm0.8$ &$3.2^{+0.6}_{-0.5}$ &$3.1\pm0.7$\\
\hline
${{\rm Fe\ K} \beta}$ & & & & & \\
Energy &$7.09^{+0.03}_{-0.04}$&- &- & -&- &-  \\
Flux &$0.43\pm0.07$ &$0.20\pm0.15$ & $0.50\pm0.20$& $0.40\pm0.15$& $0.45\pm0.10$&$0.50\pm0.15$  \\
EW & $1.1\pm0.4$&$0.4\pm0.3$ & $1.1\pm0.5$& $1.1\pm0.5$&$1.3\pm0.4$& $1.3\pm0.5$\\
\hline
${\rm Fe\ \textsc{xxv}}$ He-$\alpha$  & & & & & \\
Energy &$6.68^{+0.01}_{-0.03}$&- &- & & -& - \\
Flux & $0.32\pm0.05$& $0.20\pm0.15$&$0.30\pm0.20$ & $<0.40$&$0.30\pm0.10$&$0.50\pm0.15$ \\
EW & $0.6\pm0.3$& $0.4\pm0.3$& $0.5\pm0.3$&$<0.8$ & $0.6\pm0.3$& $1.1\pm0.3$\\
\hline
${\rm Fe\ \textsc{xxvi}}$ Ly-$\alpha$ & & & & & \\
Energy & $6.966$& -& -& -&- &- \\
Flux & $0.12\pm0.07$& $<0.15$& $0.25\pm0.15$& $<0.30$& $0.20\pm0.10$& $<0.30$\\
EW &$0.3\pm0.2$ & $<0.2$& $0.3\pm0.2$& $<0.7$& $0.4\pm0.2$&$<0.7$ \\
\hline
Compton Shoulder & $20\pm10\%$& $35^{+35}_{-20}\%$& $40^{+30}_{-20}\%$& $25\pm15\%$& $20\pm15\%$& $15^{+15}_{-10}\%$\\
& & & & & \\
F$_{\rm 3-10\ keV}$ &- & $4.7\pm0.5$& $5.6\pm0.6$& $5.0\pm0.5$&$4.8\pm0.5$&$5.1\pm0.5$ \\
\hline
\end{tabular}
\end{center}
\caption{\label{parameters} Best fit parameters. A model composed of pure reflection and five emission lines is applied to the spectra extracted from the central regions, dashes indicate fixed parameters. The \textsc{pexrav} normalization is in $10^{-3}$ photons/keV/cm$^2$/s at 1 keV, fluxes are in $10^{-5}$ photons/cm$^2$/s, equivalent widhts in keV and the Compton shoulder is calculated with respect to the narrow iron K$\alpha$ core flux. Fluxes in the 3--10 keV band are in $10^{-13}$ erg/cm$^2$/s. }
\end{table*}

\subsection{Spectroscopy}
\subsubsection{The central reprocessed emission}
\begin{figure}
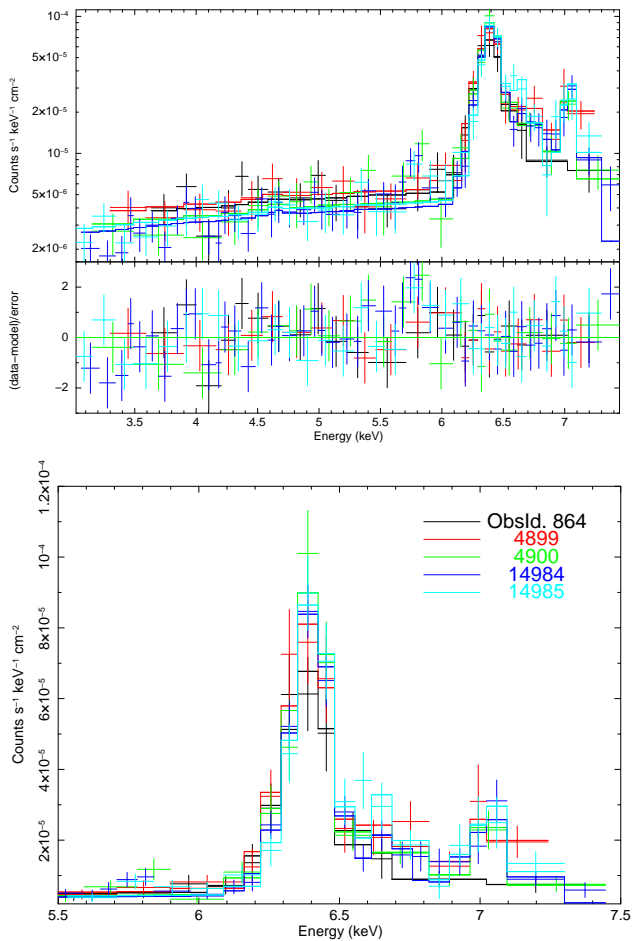

 \epsfig{file=best_fit.ps, width=0.7\columnwidth, angle=-90}
  \epsfig{file=zoom_hard.ps, width=0.78\columnwidth, angle=-90}
  \caption{{\it Top panel}:  Best fit model applied to the five spectra extracted from the nucleus and corresponding residuals are shown. Data are divided by the effective area of the instrument. {\it Bottom panel}: the Iron line complex in the five different observations is shown, in the 5.5-7.5 keV energy band.}
  \label{bestfit}
\end{figure}

We start our 3--10 keV spectral analysis by searching for any variability of the central reflected emission throughout the 13 years of monitoring, by extracting spectra from the central circular region, adopting a 1 arcsec radius, in all the five observations (``Nucleus" region in Fig. \ref{images}, top-right panel). The primary emission is reprocessed by neutral material, with a resulting Compton reflection continuum and an associated iron K$\alpha$ line \citep{gf91,mpp91}. Indeed, the five spectra were fitted with a model consisting of a reflection continuum \citep[model PEXRAV,][]{mz95} and five emission lines for neutral Fe K$\alpha$ at 6.4 keV, Fe \textsc{xxv} He-$\alpha$ (actually a triplet, with the two most intense lines, the forbidden and the resonant, at 6.637 keV and 6.700 keV), Fe \textsc{xxvi} Ly-$\alpha$ at 6.966 keV, Fe K$\beta$ at 7.058 keV and for the Compton Shoulder (CS) redwards of the line core \citep[with energy fixed at 6.3 keV and $\sigma=40$ eV, ][]{matt02}.  A value of $\Gamma=1.9$ was used for the illuminating continuum, as measured from the {\it NuSTAR} high energy spectra \citep{pcf14}, and the cosine of the inclination angle of the reflector is fixed to 0.45. A Galactic absorption of $1.38 \times 10^{21}$cm$^{-2}$ has been taken into account \citep{kalberla05}. We get a $\chi^2$/dof=113/125=0.9 and no strong residuals throughout the considered energy band ( Fig. \ref{bestfit}, top panel). The neutral iron K$\alpha$  emission line is unresolved and only an upper limit $\sigma<50$ eV is retrieved (FWHM$<$5500 km s$^{-1}$), which is in agreement with past {\it Chandra} HEG measurements \citep[FWHM=$2780^{+1110}_{-740}$ km s$^{-1}$:][]{syw11}.\\
 Best fit parameters are shown in Table \ref{parameters}: no variations among the best fit values is found, confirming the results already discussed in \citet{mrw12}. The central 1 arcsec reprocessed emission in NGC 4945 has remained constant, within a few per cent, between 2000 and 2013 (Fig. \ref{bestfit}, bottom panel).\\
The flux of the Compton Shoulder is $20\pm10 \%$ of the flux of the narrow core, consistent with expectations from Compton-thick material \citep{matt02}. The ratio between the Fe K$\beta$ and the Fe K$\alpha$ is higher than the expected value \citep[0.155--0.16:][]{mbm03}, indicating either a contribution from the Fe \textsc{xxvi} Ly-$\alpha$ emission line to the total flux of the Fe K$\beta$, or that this is moderately ionized (as suggested from its best fit energy centroid in Table \ref{parameters}). \\

\subsubsection{Spatially resolved spectroscopy}
We apply to the three pairs of spectra (coadded from observations with and without gratings, respectively), extracted from regions 1, 2 and 3, a model composed of a reflection continuum with the same parameters listed above and two narrow Gaussians (the width of the lines is fixed to zero), to reproduce the Fe K$\alpha$ and Fe \textsc{xxv} He-$\alpha$ at 6.4 keV and 6.67 keV. We fitted them separately and best fit results of the 3--10 keV spectra can be found in Table \ref{resolv_fit}. If any parameter is left free to vary between the two sets of spectra no significant differences with respect to best fit values are found. \\
The Compton reflection continuum and the associated neutral Fe K$\alpha$ emission line are constant along three selected regions (regions 1, 2 and 3), indicating an emitting material which spatially extends on scales of about 200 parsecs. The Equivalent Width of the Fe K$\alpha$ line differs from the one inferred in the central region by a factor $\sim2$, suggesting a different geometry of the circumnuclear matter in the innermost 20 parsecs. Indeed, we measured a constant Fe K$\alpha$ EW=$2.7\pm0.5$ keV in the spectra extracted from the central 1 arcsec circular region. When, on the other hand, spectra extracted from regions 1, 2 and 3 are considered, the EW  is significantly lower (EW=0.45--0.75 keV). When spectra from region 4 are considered (the southern clump of intense neutral iron emission in Fig. 3, bottom-left panel) we retrieve an Fe K$\alpha$ EW=$2.2^{+1.3}_{-0.9}$ keV, consistent with the nuclear one.\\
The Fe \textsc{xxv} He-$\alpha$ flux is significantly higher in region 1 (3.2$\sigma$ and 3.5$\sigma$ than region 2 and 3, respectively).  Spectra and best fit models are shown in Fig. \ref{bestfit_resolv} (top panels) while, on the bottom panel, we show the contour plots between the normalization and energy centroids in the best fit of the spectra extracted from Region 1. \\
Some additional soft emission component is required to model the spectra from Region 3. This can be ascribed to superwind-starburst-driven, metal-enriched galactic-scale outflows \citep{vcb05} that are dominated by emission from hot plasma, as shown by \citet{srw02}. A detailed investigation is beyond the scope of this work but the inclusion of a simple thermal component (\textsc{mekal} model in \textsc{Xspec}) gives a gas temperature kT=$2.0_{-1.0}^{+1.5}$ keV with a contribution of $\sim15\%$ to the total 3--10 kV flux, in agreement with \citet{srw02} and \citet{pcf14}. \\

\section{Discussion}
\subsection{Equivalent width of the Iron K$\alpha$ line}
In our previous paper \citep{mrw12}, we inferred a distance of the reflecting structure D$>$35~pc, by comparing the very variable high energy {\it Swift}-BAT light curve and the $4\%$ constancy of the reflected emission below 10 keV, on a time scale of ten years. The additional 200 ks ACIS-S data allowed us to picture with even more detail the central structure of this source and to discover an enhanced iron emission in the innermost nuclear region, with respect to the associated Compton reflection continuum.  \\
Strongly absorbed Seyfert galaxies are the perfect targets for studying the properties of the cold reprocessor, because the Compton reflection features are not diluted by the presence of a strong primary continuum. Indeed, \citet{yws01} firstly showed that the neutral iron K$\alpha$ emitter is extended in NGC 1068. Such a material is well imaged by the ACIS-S and extends up to $\sim$20 arcsec to the northeast and southwest which, at the redshift of the source, corresponding to about 2.2 kpc. \citet{baw15} recently confirmed this result, using {\it NuSTAR}, by spectrally discriminating three neutral reflectors, which differently contribute to the iron K$\alpha$ and Compton hump emission. Other Compton-thick sources have been used to map the extended iron K$\alpha$ emission line: Mrk 3 \citep{glm12, gra16}, Circinus \citep{mmb13} and they all indicate that reflecting material may extend on scales ranging from tens to thousands of parsecs. In particular, the last source showed a spatially variable Equivalent  Width of the iron K$\alpha$ which strongly resembles the one that we have presented for NGC 4945. However, while the ratio between the line and the associated reflection continuum is axisymmetric around the nucleus in Circinus, we show (Fig. \ref{images}, bottom-left panel) that this is not the case for NGC 4945. The iron K$\alpha$ emission is enhanced towards the nucleus (EW=$2.7\pm0.5$ keV) and in a southern clump (EW$_{\rm reg 4}$=$2.2^{+1.3}_{-0.9}$ keV) with respect to the other circumnuclear regions (EW$_{\rm reg 1}$=$0.45^{+0.30}_{-0.20}$ keV, EW$_{\rm reg 2}$=$0.65^{+0.30}_{-0.25}$ keV and EW$_{\rm reg 3}$=$0.75^{+0.40}_{-0.25}$ keV), with no statistically significant differences between the last three regions.\\

\begin{table*}
\begin{center}
\begin{tabular}{ccccc}
{\bf Parameter} & {\bf Reg. 1} & {\bf Reg. 2} & {\bf Reg. 3 }& {\bf Reg. 4 } \\
\hline
N$_{\rm pexrav}$ & $0.40\pm0.03$& $0.39\pm0.03$ &$0.37\pm0.08$ & $0.12\pm0.02$\\
\hline
${{\rm Fe\ K} \alpha}$ & &   & &\\
Energy&$6.44\pm0.05$&$6.43\pm0.03$ & $6.40\pm0.03$ &$6.40^{+0.02}_{-0.03}$\\
Flux &$0.05\pm0.02$ & $0.08\pm0.02$& $0.09\pm0.03$ &$0.07\pm0.02$\\
EW & $0.45^{+0.30}_{-0.20}$&$0.65^{+0.30}_{-0.25}$ &  $0.75^{+0.40}_{-0.25}$& $2.15^{+1.30}_{-0.85}$\\
\hline
${\rm Fe\ \textsc{xxv}}$ He-$\alpha$& &   & &\\
Energy&$6.65^{+0.03}_{-0.04}$& $6.66\pm0.07$&$6.65\pm0.06$ & $6.60\pm0.10$\\
Flux &$0.11\pm0.03$ & $0.04\pm0.02$&$0.03\pm0.02$ & $0.02\pm0.01$\\
EW & $0.90\pm0.30$& $0.30\pm0.25$& $0.35\pm0.30$ &$0.60^{+0.70}_{-0.45}$\\
 & &   & \\
F$_{\rm 3-10\ keV}$ &$0.80\pm0.07$ & $0.75\pm0.08$  & $0.85\pm0.08$& $0.28\pm0.05$\\
\hline
C/d.o.f. & 37/51& 65/40& 28/38 & 26/31\\
\hline
\end{tabular}
\end{center}
\caption{\label{resolv_fit} Best fit parameters to the spectra extracted from the circumnuclear regions. Same units as Table \ref{parameters}.}
\end{table*}

\begin{figure*}
 \epsfig{file=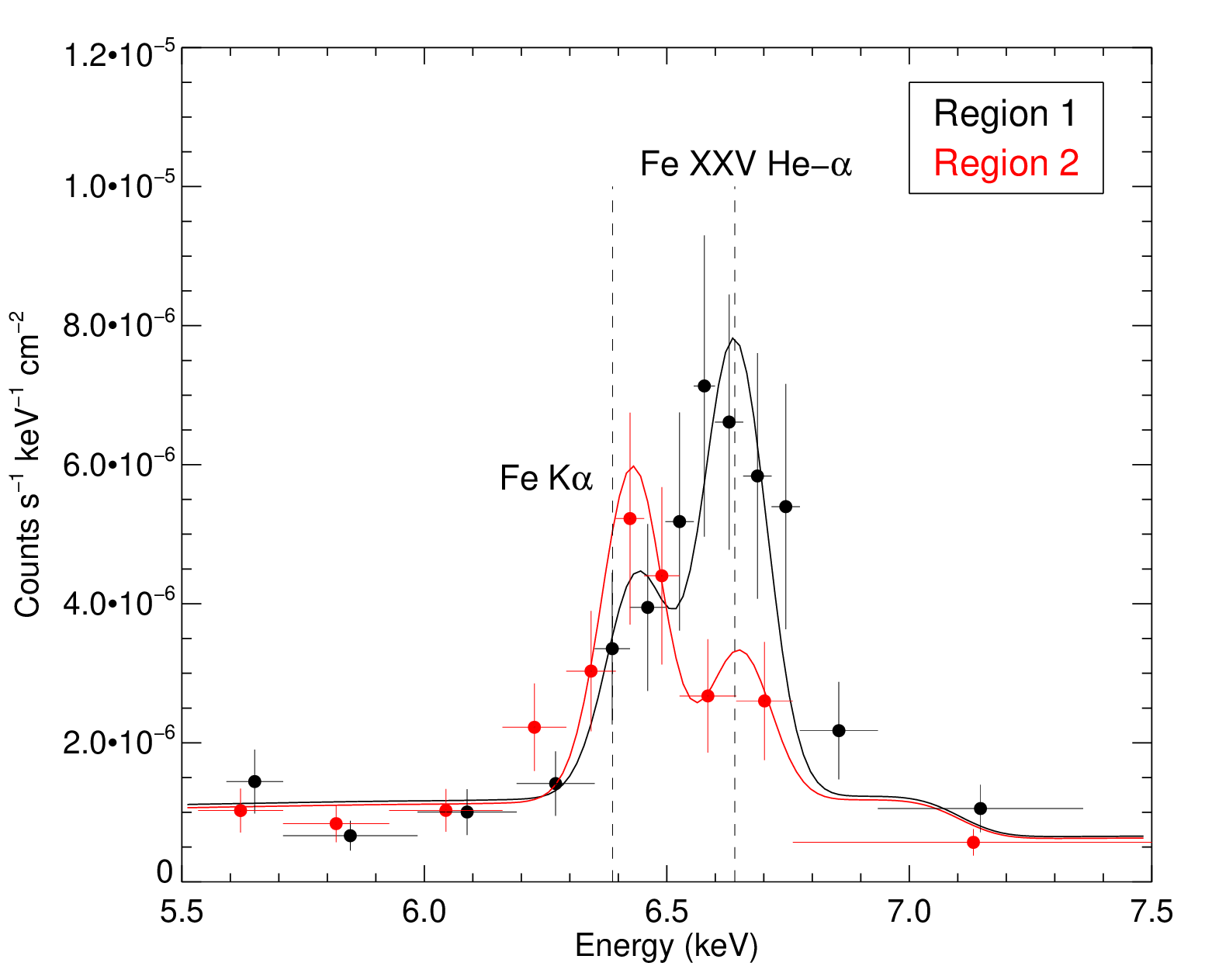, width=\columnwidth}
  \epsfig{file=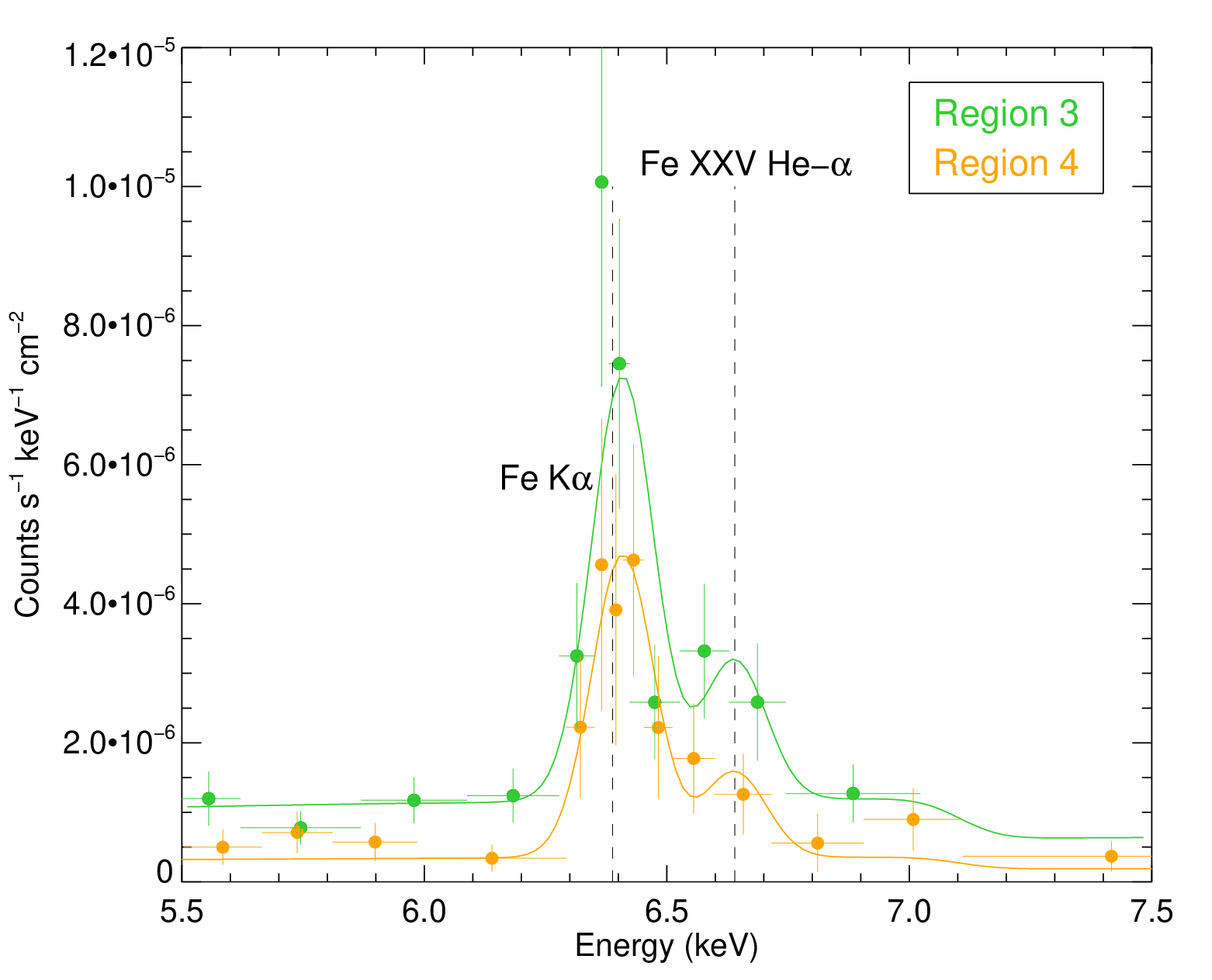, width=\columnwidth}
 \epsfig{file=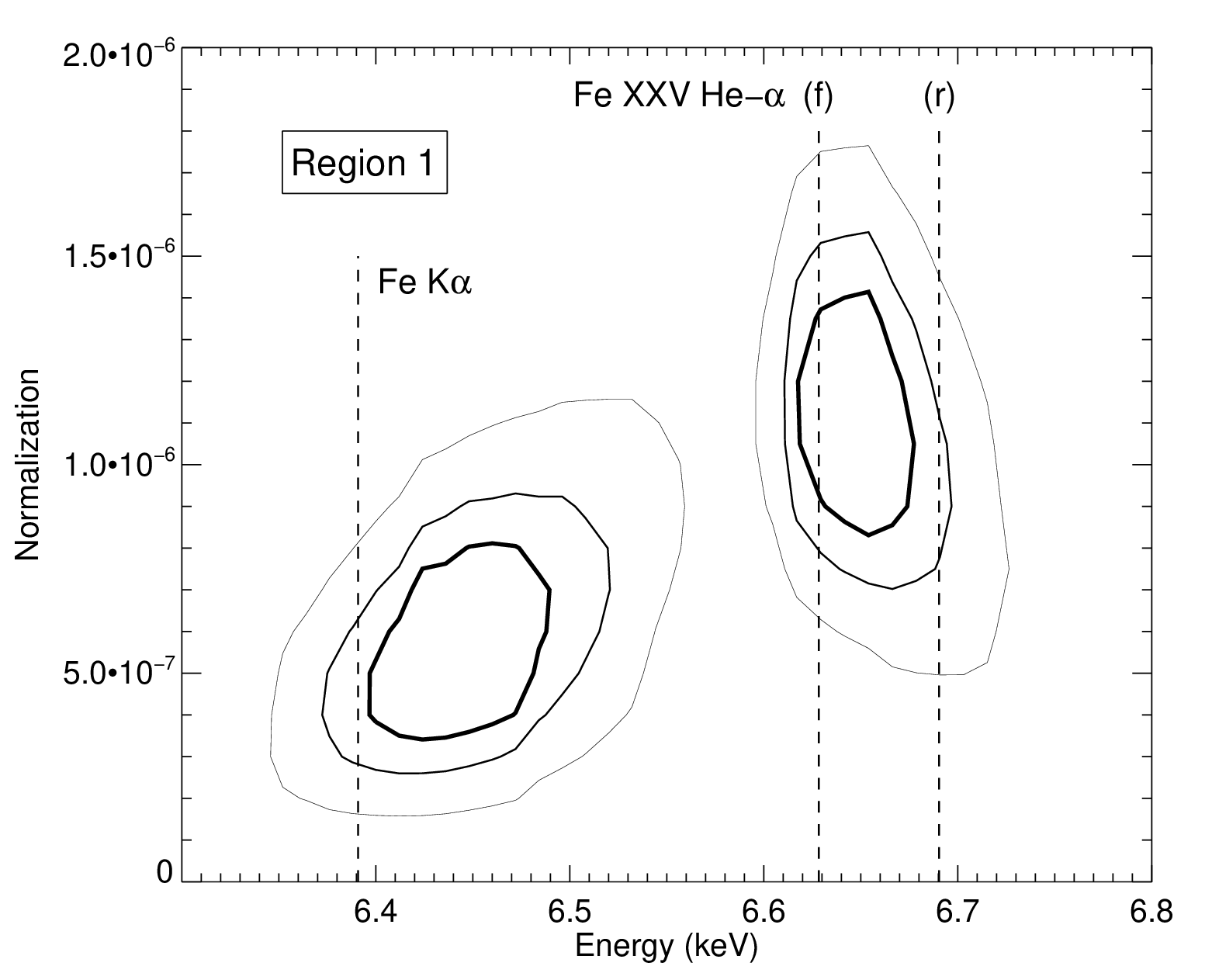, width=\columnwidth}
  \caption{\textit{Top panels:} Coadded spectra from ObsIDs 864, 14984 and 14985 obtained from regions 1, 2 3 and 4 and best fit models (see text for details). Dashed lines indicate neutral Fe K$\alpha$ and Fe \textsc{xxv} He-$\alpha$. Data from ObsIDs 4899 and 4900 are not shown for the sake of visual clarity. \textit{Bottom panel:} Contour plots between normalization of the two Gaussians and energy centroids in Region 1. Thinner to thicker solid contours indicate 99\%, 90\% and 68\% confidence levels. }
  \label{bestfit_resolv}
\end{figure*}

\subsubsection{Geometrical effects}
\begin{figure}
 \epsfig{file=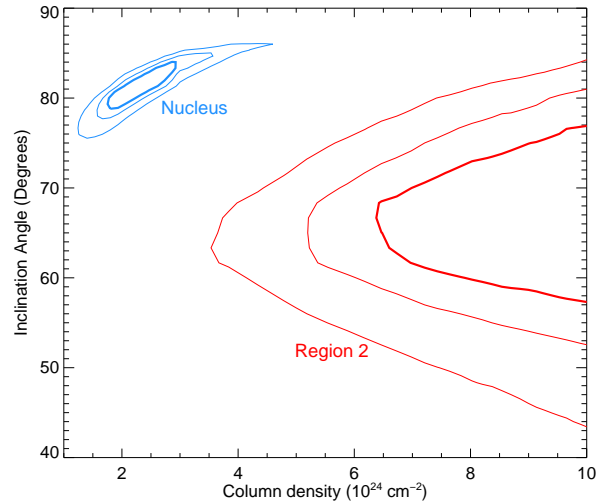, width=1.03\columnwidth}
  \caption{Contour plots between the inclination angle and the column density of the cold reflectors in the nuclear region and in region 2 (blue and red, respectively). Thinner to thicker solid contours indicate 99\%, 90\% and 68\% confidence levels. }
  \label{myt}
\end{figure}

The Fe K$\alpha$ EW, depends on the iron abundance \citep{mfr97}, on the angle $\theta_i$ between the polar direction and the line of sight \citep{mpp91,gf91} and on the column density of the illuminated material, as discussed in several works using either a toroidal \citep{yaqoob10} or a slab \citep{matt02} geometry. \\
 We chose the spectrum of one of the three circumnuclear regions with a low Fe K$\alpha$ EW (region 2) and compared it with the highest S/N spectrum from the nucleus.
Let us first assume that the changes in the Fe K$\alpha$ EW is due to the iron abundance. 
Using the model \textsc{pexmon} in \textsc{xspec}, we inferred iron abundances A(Fe)$_{\rm nucleus}=3.2\pm0.4$ and A(Fe)$_ {\rm reg 2}=0.8\pm0.5$ (with respect to solar values) for the two regions, with no significant change to other parameters. As already discussed in \citet{mmb13} for the Circinus galaxy, these large variations are physically hard to explain on such small spatial scales (about 40 parsecs).\\
To test if the changes in EW can be ascribed to different column densities or inclination angles, we used the \textsc{mytorus} model to reproduce the 2013 nuclear spectra (the ones with the highest number of counts) and the ones from region 2. We linked the normalizations, inclination angles, photon indices and column densities between the tables reproducing both the scattered and line components. In this way, we assume that the material responsible for the Compton reflection and iron K$\alpha$ emission line is one and the same. We find a good fit %($\chi^2$/dof=90/79=1.1) 
with best fit parameters N$_H=2.0^{+1.0}_{-0.3}\times10^{24}$ cm$^{-2}$, $\theta_i=80^{+4}_{-2}$ deg for the nucleus and N$_H>4\times10^{24}$ cm$^{-2}$, $\theta_i=60\pm12$ deg for region 2. Contour plots between the two parameters can be seen in Fig. \ref{myt}. Obviously, the low quality of the spectra and the lack of a high energy spectral coverage do not allow us to give strong constraints on the column density and inclination angle of the toroidal reprocessor. However, we showed that the observed EWs can be recovered only in terms of variations of N$_H$ and $\theta_i$ of the reflector. 
\subsubsection{Interplay between different continua}
Another possible explanation for the spatially variable Fe K$\alpha$ EW is the interplay between the underlying continua, associated to the neutral and ionized reflecting components. Many Seyfert 2 galaxies show intense ionized Fe emission lines in their spectra \citep{bianchi05}. The archetypal Compton-thick source NGC 1068, for example, clearly shows an intense warm reflector in addition to the cold one \citep{ifm97,matt04,pv06,baw15}: the EW of the Fe lines in the 6-7 keV band can be therefore calculated with respect to the associated continuum. In our analysis, the EWs are inferred with respect to the same Compton reflection continuum (modeled with \textsc{pexrav}) and the quality of the spectra does not allow us to separate the neutral component from an additional, ionized one. If we assume that the EW of the neutral Fe K$\alpha$  is constant within all the circumnuclear regions (EW$=2.5$ keV, with respect to its Compton reflection continuum), its decrease could be ascribed to an increase of the ionized reflection continuum, associated to the Fe \textsc{xxv} He-$\alpha$ and  Fe \textsc{xxvi} Ly-$\alpha$ emission lines. Within the given uncertainties, this scenario is consistent with our findings. When the EW of the neutral Fe K$\alpha$ decreases ($\sim0.5$ keV) the Fe \textsc{xxv} He-$\alpha$ emission lines is always detected (region 2 and 3) and it becomes even dominant in region 1: the warm reflector continuum would then prevail ($\sim60\%$ of the total flux in the 3--10 keV band) over the cold, neutral one.   
\subsection{Diffuse highly-ionized Fe emission}
  Fe \textsc{xxv} He-$\alpha$ and  Fe \textsc{xxvi} Ly-$\alpha$ emission lines can be produced via recombination and resonant scattering \citep[][and references therein]{mbf96,bm02,bianchi05}. In particular, the ratio of the He-like Fe triplet lines (forbidden, intercombination and resonant) can tell us about the state of the emitting gas. In a photoionized gas, emission from the resonant $r$ iron line is weak, while the forbidden $f$ and the intercombination $i$ lines contribute almost equally to the total emission, considering a wide range of densities \citep{pd00, bk00}. On the other hand, in a collisionally ionized gas, the ratio of the lines points in favor of a dominant resonant line, with respect to the forbidden and intercombination lines, such as the case of NGC 6240 \citep{bkh03, wnf14}.  In NGC 4945, the  Fe \textsc{xxv} He-$\alpha$ complex is dominated by the forbidden $f$ line (Fig. \ref{bestfit_resolv}, bottom panel), with only an upper limit to the resonant $r$ component when the two energy centroids are fixed. This is in favor of a scenario in which the clump of Fe \textsc{xxv} He-$\alpha$ gas pictured in Fig. 1 and 3 is photoionized by the nuclear continuum. In this photoionization configuration, the cross-section of the resonant scattering is much larger than the recombination one and the gas becomes rapidly thick to the former process. Even though recombination is less dominant than resonant scattering in the optically thin regime, it has been shown that column densities as small as $5\times10^{20}$ cm$^{-2}$ can significantly suppress the EW of emission lines produced by resonant scattering \citep[][]{mbf96}. As a consequence, the resonant $r$ emission line dominates the spectrum up to $\simeq10^{22}$ cm$^{-2}$, then it is reached and overcome by the forbidden $f$ and intercombination $i$ lines, which become more intense at higher column densities. \\
 %Using the two measured values (F$_{\rm (f)}=0.9_{-0.3}^{+0.1}\times10^{-6}$ photons/cm$^2$/s and F$_{\rm (r)}<0.8\times10^{-6}$ photons/cm$^2$/s) and figure 7 from \citet[][which summarizes the relative intensity of the $f$, $r$ and $i$ lines versus the column density of the emitting gas]{bianchi05} we are able to give a rough constraint of log[N$_{\rm H}({\rm cm}^{-2})]>22.5$. \\
 To verify our conclusion and measure the dependence from the Fe abundance and ionization state, we produced a grid model for \textsc{xspec} using \textsc{cloudy} 13.03 \citep[last described by][]{cloudy}. It is an extension of the same model used in \citet{bianchi10} and \citet{mbm11}. The main ingredients are: plane parallel geometry, with the flux of photons striking the illuminated face of the cloud given in terms of ionization parameter $U$ \citep{of06}; incident continuum modeled as in \citet{korista97}; constant electron density $\mathrm{n_e}=10^5$ cm$^{-3}$; elemental abundances as in Table 7.5 of \textsc{cloudy} documentation\footnote{\sloppy Hazy 1 version 13, p. 65: \url{http://viewvc.nublado.org/index.cgi/tags/release/c13.05/docs/hazy1_c13.pdf?revision=11558&root=cloudy}}; grid parameters are $\log U=[1.00:4.00]$, step 0.25, $\log N_\mathrm{H}=[23.0:25.0]$, step 0.1, and $A_{Fe}=[1.0:5.0]$. Only the reflected spectrum, arising from the illuminated face of the cloud, has been taken into account in our model. We also produced tables with different densities ($\mathrm{n_e}=10^3-10^4$ cm$^{-3}$): all the fits in this paper resulted insensitive to this parameter, as expected since we are always treating density regimes where line ratios of He-like triplets are insensitive to density \citep{bk00,pd00}. When this model is applied to the spectra extracted from Region 1 we find $\log U=2.2^{+0.3}_{-0.2}$, $\log N_\mathrm{H}>23.5$ and $A_{Fe}>3$ with no significant variations to the neutral Fe K$\alpha$ parameters found in Sect. 3.2.2. The adimensional ionization parameter $U$ is defined as:
 \begin{eqnarray}
  U=\frac{\int^{\infty}_{\nu_{\rm R}} \frac{L_\nu}{h\nu} d\nu}{4\pi c r^2 n},
 \end{eqnarray}
where $c$ is the speed of light, $r$ the distance of the gas from the illuminating source (in our case $r\sim 40$ pc), $n$ its density and $\nu_{\rm R}$ the frequency corresponding to 1 Rydberg. By inverting the previous equation and using a 2--10 keV luminosity L$_{\rm X}=4\times 10^{42}$ erg s$^{-1}$ \citep{pcf14}, we can derive a density $n_{cl}\simeq10^2$ cm$^{-3}$ for the clump. 
\subsection{Clumpiness of the reflector}
The complex morphology at the center of NGC 4945 suggests, once again, that the circumnuclear environment of heavily obscured sources is not homogeneous and the absorbing/reflecting structures are multiple and distributed on different spatial scales \citep{bmr12}. Since the seminal works about clumpy tori \citep{nsn08} there is growing evidence against the monolithic obscurer envisaged by unified models \citep{baw15,mcb16,mbm16} and the presence of a clump of ionized Fe in NGC 4945 (with a different density compared to its surroundings) supports this physical scenario.\\
Thanks to the imaging analysis presented in Sect. 3.1 we are able to give rough constraints on the number and size of the Compton-thick clouds in the circumnuclear environment of NGC 4945. Fig. \ref{subpix_images} shows that the reflected emission above 3 keV is homogeneously distributed around the nucleus, on spatial scales of approximately 250 pc$\times$150 pc on the equatorial plane. On the other hand, the covering factor of such a reflecting structure is constrained to be very small \citep[$<10\%$:][]{mzd00,dmz03,yaq12,pcf14}, to allow the variable primary continuum to pierce through the absorber. \\
The angular resolution of {\it Chandra} is $\sim0.2$ arcsec, while the pixel size is 0.495 arcsec. We can estimate the number of clouds that we are not able to resolve within a 5 arcsec radius circular region centered on the nucleus. We obtain a circle with a total projected area of 78.5 arcsec$^2$ and we need at least N$_{cl}>100$ to have a homogeneous, unresolved structure. Each of these clumps will have a radius $r\simeq9$ pc and if we then assume that each Compton thick cloud has a covering factor of $5\%$ the area of such inferred clumps, we obtain an upper limit on the radius of the clouds $r_{cl}<2$ pc.
%begin{eqnarray}
%Ac=5\%A
%end{eqnarray}

\section{Conclusions}
We presented the imaging and spectroscopic analysis of a combined, 420 ks long, {\it Chandra} observation of the Compton-thick galaxy NGC 4945. The main results of this paper can be summarized as follows.
\begin{enumerate}
 \item The neutral iron K$\alpha$ emission line and the associated Compton reflection continuum are extended and can be spatially resolved, on scales of hundreds of parsecs. The neutral iron emission  is enhanced in the central, unresolved, region and the spectral analysis reveals a spatial variation in the EW of the K$\alpha$ line with respect to the circumnuclear regions;\\
 \item a clump of gas with an intense Fe \textsc{xxv} He-$\alpha$ emission (3.2$\sigma$ and 3.5$\sigma$ higher than the two other blobs considered) is detected, at a distance $d\sim 40$ pc from the nucleus. The dominant transition of the Fe \textsc{xxv} He-$\alpha$ triplet is the forbidden one, suggesting a scenario in which the gas is photoionized from the primary continuum of the AGN. When the spectrum extracted from this region is fitted using a self-consistent photoionization code, a density $n_{\rm cl}\simeq10^2$ cm$^{-3}$ is inferred.
 \end{enumerate}
 Our findings support a physical environment around the nucleus which is not homogeneously distributed: the gas responsible for the reprocessing of the nuclear radiation is clumpy and  extended around the central region.

\section*{ACKNOWLEDGEMENTS}
The authors thank the anonymous referee for her/his comments and suggestions that improved the manuscript.
AM, SB and GM acknowledge financial support the European Union Seventh Framework Programme (FP7/2007-2013) under grant agreement n.312789. EN acknowledges funding from the European Union's Horizon 2020 research and innovation programme under the Marie Sk\l{l}odowska-Curie grant agreement No. 664931. JW acknowledges support from National Key Program for Science and Technology Research and Development 2016YFA0400702, and the NSFC grants 11473021, 11522323.

\bibliographystyle{mn2e}
\bibliographystyle{mn2e}
\bibliography{sbs} 

\end{document}